\pdfoutput=1
\documentclass[doublecol]{epl2} 
\usepackage{color}
\usepackage{graphicx,amsmath}
\newcommand{\nn}{\nonumber}

\newcommand{\be}{\begin{equation}}
\newcommand{\ee}{\end{equation}}
\newcommand{\bea}{\begin{eqnarray}}
\newcommand{\eea}{\end{eqnarray}}

\newlength{\figsize}
\newcommand{\cm}[1]{}

\begin{document}

\title{Directed polymer near a hard wall and KPZ equation in the half-space}
\author{Thomas Gueudre and Pierre Le Doussal} 
\institute{CNRS-Laboratoire
de Physique Th{\'e}orique de l'Ecole Normale Sup{\'e}rieure, 24 rue
Lhomond,75231 Cedex 05, Paris, France} 
\date{\today\ -- \jobname}
\abstract{We study the directed polymer with fixed endpoints near an absorbing wall, in the continuum and in presence of disorder, equivalent to the KPZ equation on the half space with droplet initial conditions.
From a Bethe Ansatz solution of the equivalent attractive boson model we obtain the exact expression for
the free energy distribution at all times. It converges at large time to the Tracy Widom distribution $F_4$ of the
Gaussian Symplectic Ensemble (GSE). We compare our results with numerical simulations of the lattice
directed polymer, both at zero and high temperature.}

\maketitle

Much progress was achieved recently in finding exact solutions in one dimension for noisy growth models
in the Kardar-Parisi-Zhang (KPZ) universality class \cite{KPZ,kpzreviews}, and for the closely related 
equilibrium statistical mechanics problem of the directed polymer (DP) in presence of quenched disorder \cite{directedpoly}.
The KPZ class has been explored in several recent experiments \cite{exp4,exp5},
and the DP has found applications ranging from biophysics \cite{hwa} to describing the glass phase 
of pinned vortex lines \cite{vortex} and magnetic walls \cite{lemerle}.
The height of the growing interface, $h(x,t)$, corresponds to the free energy of a DP of length $t$
starting at point $x$, under a mapping which is exact in the continuum (Cole-Hopf), as
well as for some discrete realizations. Not only the scaling exponents $h \sim t^{1/3}$, $x \sim t^{2/3}$ are known 
\cite{exponent,Johansson2000}, 
but also the one-point (and in some cases the many-point) probability distribution (PDF)
of the height have been obtained \cite{reviewCorwin,spohnreview}. Their dependence in the initial condition was found to
exhibit remarkable universality at large time, with only a few subclasses, most being related to
Tracy Widom (TW) distributions \cite{TW1994} of largest eigenvalues of random matrices. Most of these 
subclasses were initially discovered in a discrete growth model (the PNG model)
\cite{spohn2000,pngsources,ferrari1} which can be mapped onto the
statistics of random permutations  \cite{randompermutations}, and a 
zero temperature lattice DP model \cite{Johansson2000}.
Recently, exact solutions have been obtained directly in the continuum at arbitrary time $t$,
for the droplet \cite{spohnKPZEdge,we,dotsenko,corwinDP}, flat \cite{we-flat,we-flatlong} 
and stationary \cite{SasamotoStationary} initial conditions. The PDF of the height $h(x,t)$ 
converges at large time to $F_2$, the Gaussian unitary ensemble (GUE), and to $F_1$, the Gaussian orthogonal ensemble
(GOE) universal TW distributions, for droplet and flat initial conditions respectively. 
One useful method which led to these solutions introduces $n$ replica and maps
the DP problem to the Lieb Liniger model, i.e. the quantum mechanics of $n$ bosons with mutual delta-function
attraction, a model which can be solved using the Bethe Ansatz. 

The KPZ equation on the half line $x>0$, equivalently a DP in presence of a wall, is also
of great interest. In the statistical mechanics context constrained fluctuations
are important for the study of fluctuation-induced (Casimir) forces \cite{casimir,kaycasimir}  and
for extreme value statistics. In the surface growth context one can study
an interface pinned at a point, or an average growth rate which jumps 
across a boundary. The half space problem was studied previously in a
discrete version, for the (symmetrized) random permutations/PNG model 
\cite{BaikSymPermutations,sasamotohalfspace} and 
found to also involve TW distributions in the limit of large system size.
In order to exhibit full KPZ universality, it is important to solve the problem directly in the continuum, i.e. for the
KPZ equation itself. Furthermore, previous approaches did not provide any information about the finite 
time behavior which is also universal \cite{footnote00}.
%
%

\begin{figure}[h]
\centering
\includegraphics[scale=1.1]{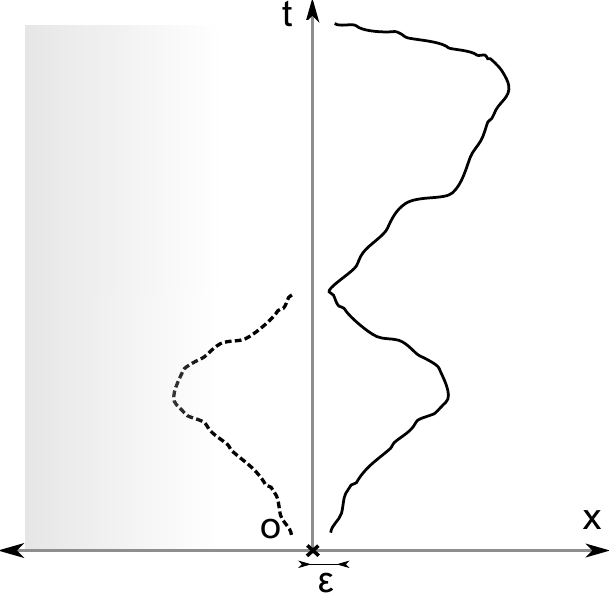}
\caption{Solid line: a DP with both endpoints fixed at small $x=\epsilon$ with a hard wall 
at $x=0$: the DP probability vanishes at the wall. Dashed line: mirror image discussed at the end.}
\label{f:dp}
\end{figure}

The aim of this Letter is to present a solution of the directed polymer problem in the continuum in presence of
a hard wall (absorbing wall) using the Bethe ansatz (BA). Equivalently, we obtain the
one-point height probability distribution for the KPZ equation on the half line $x>0$ with
fixed large negative value of $h$ or of $- \nabla h$ (i.e. a small contact angle) at $x=0$. For simplicity we study a DP with both endpoints fixed - which corresponds to
the droplet initial condition in KPZ - near the wall. We do not consider
the case of the attractive wall although we briefly
mention it at the end. We obtain an exact expression for the generating function
of the moments of the DP partition sum as a Fredholm Pfaffian, from 
which we extract the PDF of the free energy of the DP 
(height of KPZ) at all times. We then show that this PDF 
converges to $F_4$, the Tracy Widom distribution of the
largest eigenvalue of the Gaussian Symplectic Ensemble (GSE).
The calculation is performed on the DP formulation, the consequences
for the KPZ equation being detailed at the end.
Our results are checked against numerics on a discrete DP model,
both at high and zero temperature, thereby confirming universality. 
Some consequences for extreme value
statistics are discussed. Note that this is the first occurrence of the $F_4$ distribution
and of the GSE within a continuum BA calculation. It is
consistent with the results of \cite{BaikSymPermutations,sasamotohalfspace} for the discrete model
and confirms that these belong to the same universality class than the continuum KPZ equation
on the half space, solved here for all times. 

{\it Directed polymer: analytical solution.} We consider the partition function of a DP at temperature $T$ in the continuum,
i.e the sum over positive paths $x(\tau) \in R^+$ starting at $x(0)=y$ and ending at $x(t)=x$
\be \label{zdef} 
Z(x,y,t) = \int_{x(0)=y}^{x(t)=x}  Dx(\tau) e^{- \frac{1}{T} \int_0^t d\tau [ \frac{1}{2}  (\frac{d x}{d\tau})^2  + V(x(\tau),\tau) ]} \,,
\ee
with initial condition $Z(x,y,0)=\delta(x-y)$. The hard wall is implemented by requiring that $Z(0,y,t)=Z(x,0,t)=0$.
The random potential $V(x,t)$ is centered gaussian with correlator $\overline{V(x,t) V(x',t)} = \bar c \delta(t-t') \delta(x-x')$.
The natural units for the continuum model are $t^*= 2 T^5/\bar c^2$ and $x^*=T^3/\bar c$ which allow to remove $T$ and
set $\bar c=1$ \footnote{in final result performing $x \to x/x^*$, $t \to t/t^*$ and in the free energy $F=-T \ln Z$,
restores $T$ dependence.} The time (i.e. polymer length) dependence is embedded in
a single dimensionless parameter:
\bea \label{lambdadef}
\lambda = (t/4 t^*)^{1/3} 
\eea 
as defined in our previous works \cite{we,we-flat,we-flatlong} and in \cite{dotsenko}.

Replicating (\ref{zdef}) and averaging over disorder one finds \cite{kardareplica} that 
the $n$-th integer moment of the DP partition sum can be expressed as a quantum mechanical 
expectation for $n$ particles described by the attractive Lieb-Liniger Hamiltonian \cite{ll}
\be
H_n = -\sum_{j=1}^n \frac{\partial^2}{\partial {x_j^2}} 
- 2 \bar c \sum_{1 \leq  i<j \leq n} \delta(x_i - x_j).
\label{LL}
\ee
in natural units (for the moment not rescaling by $\bar c$, as in \cite{we}).
The moments of the partition sum with both endpoints fixed at $x$ can 
be written as:
\bea \label{sumstates}
\overline{Z(x,x,t)^n} = \sum_\mu |\Psi_\mu(x,..,x)|^2 \frac{e^{- t E_\mu} }{||\mu||^2} 
\eea 
i.e. a sum over the un-normalized eigenfunctions $\Psi_\mu$ (of norm denoted $||\mu ||$) 
of $H_{n}$ with energies $E_\mu$. Here we used the fact that only symmetric (i.e. bosonic) eigenstates contribute.
In presence of a hard wall at $x=0$ we must impose that $\Psi_\mu(x_1,..,x_n)$ vanishes when any of the $x_j$ vanishes.
This case can also be solved, by a simple generalization of the standard BA \cite{BAhardwall,ChineseBA}. The Bethe states  $\Psi_\mu$ are 
superpositions of plane waves \cite{ll}  over all permutations $P$ of the rapidities $\lambda_j$ ($j=1,..n$), with here an
additional summation over $\pm \lambda_j$. The eigenfunctions read, for $x_1<..<x_n$:
\bea \label{wave}
&& \Psi_\mu(x_1,..,x_n) =  \frac{1}{(2 i)^{n-1}}  \sum_{P \in S_n} \sum_{\epsilon_2,..\epsilon_n=\pm 1} \epsilon_2..\epsilon_n \\
&& \times A_{\lambda_{P_1},\epsilon_2 \lambda_{P_2}, ..,\epsilon_n \lambda_{P_n}}  \sin(x_1 \lambda_1) \prod_{j=2}^n e^{i \epsilon_j x_j \lambda_{P_j} } \\
&& A_{\lambda_1,..,\lambda_n}= \prod_{n \geq \ell > k \geq 1} (1+ \frac{i \bar c}{\lambda_\ell - \lambda_k})(1+ \frac{i \bar c}{\lambda_\ell + \lambda_k})
\eea
recalling that $\Psi_\mu(x_1,..x_n)$ is symmetric in its arguments.
Imposing a second boundary condition at $x=L$, e.g. also a hard wall, one gets the corresponding
Bethe equations \cite{BAhardwall} which determine the possible sets of $\lambda_j$. 
The large $L$ limit was studied in \cite{ChineseBA} and we do not reproduce the analysis
here. The structure of the states is found very similar to the standard case, i.e. the 
general eigenstates are built by partitioning the $n$ particles into a set of $n_s$ 
bound-states formed by $m_j \geq 1$ particles with $n=\sum_{j=1}^{n_s} m_j$.
Each bound state is a {\it perfect string} \cite{m-65} , i.e. a set of
rapidities $\lambda^{j, a}=k_j +\frac{i\bar c}2(m_j+1-2a)$, where $a = 1,...,m_j$ 
labels the rapidities within the string. Such eigenstates have momentum 
$K_\mu=\sum_{j=1}^{n_s} m_j k_j$ 
and energy $E_\mu=\sum_{j=1}^{n_s} (m_j k_j^2-\frac{\bar c^2}{12} m_j(m_j^2-1))$. 
The difference
with the standard case is that the states are now invariant by a sign change of any
of the momenta $\lambda_j \to - \lambda_j$, i.e. $k_j \to -k_j$.

To simplify the problem, we restrict here to a DP with endpoints near the wall, i.e
we define the partition sum for $x=\epsilon=0^+$ (see Fig. \ref{f:dp}) :
\bea
Z = \lim_{x \to 0^+} Z_V(x,x,t)/x^2 
\eea 
Then the factor involving the wave function in (\ref{sumstates}) drastically simplifies as 
$\lim_{x \to 0^+}  |\Psi_\mu(x,..x)|^2/x^{2n}= n!^2 \lambda_1^2..\lambda_n^2$. The last needed
factor in (\ref{sumstates}) is the norm, usually not trivial to obtain \cite{cc-07}. With some amount of heuristics
we arrive at the following formula \cite{us-long} (we now fully use the natural units, hence setting $\bar c=1$):
\bea
&& ||\mu||^2 =  n!  2^{-n_s}
\prod_{i=1}^{n_s} S_{k_i,m_i} \prod_{1 \leq i<j \leq n_s} D_{k_i,m_i,k_j,m_j}  L^{n_s} \nn \\
&& S_{k,m} =  \frac{m^2}{2^{2m-2}}  \prod_{p=1}^{[m/2]} \frac{k^2+(m+1-2 p)^2/4}{k^2 + (m-2 p)^2/4} 
\eea 
and
\bea
&& D_{k_1,m_1,k_2,m_2} =
\frac{4 (k_1-k_2)^2 + (m_1+m_2)^2 }{4 (k_1-k_2)^2 + (m_1-m_2)^2 } \nn \\
&& \times \frac{4 (k_1+k_2)^2 + (m_1+m_2)^2 }{4 (k_1+k_2)^2 + (m_1-m_2)^2 } 
\eea
We now have a starting formula for the integer moments 
\bea \label{partsum}
&& \overline{Z^n} = \sum_{n_s=1}^n \frac{  n! 2^{n_s} }{n_s! }  \sum_{(m_1,..,m_{n_s})_n} \\
&& 
\prod_{j=1}^{n_s}  \int \frac{d k_j}{2 \pi}  \frac{b_{k_j,m_j}}{4 m_j} e^{ (m_j^3-m_j) \frac{ t}{12} - m_j k_j^2 t } 
 \prod_{1 \leq i<j \leq n_s}  D_{k_i,m_i,k_j,m_j} \nn 
 \eea
with $b_{k,m} =  \prod_{j=0}^{m-1} (4 k^2 + j^2 )$.
Here $(m_1,..,m_{n_s})_n$ stands for all the partitioning of $n$ such that $\sum_{j=1}^{n_s} m_j=n$ with $m_j \geq 1$
and we used $\sum_{k_j} \to m_j L \int \frac{dk}{2\pi}$ which holds also here in the large $L$ limit.

This formula allows for predictions at small time. Defining \cite{footnote0} $z = Z/\overline{Z}$ 
we obtain $\overline{z^2}^c = \overline{z^2} -1$ as:
\bea
&& \overline{z^2}^c = \sqrt{\frac{\pi }{2}} e^{2 \lambda ^3} \left(4 \lambda
   ^3+3\right) \lambda ^{3/2} (\text{erf}(\sqrt{2}
   \lambda ^{3/2})+1)+2 \lambda ^3 \nn \\
&& = \frac{3}{2} \sqrt{2 \pi} \lambda ^{3/2}+8 \lambda ^3+O(\lambda ^{9/2})
\eea 
and, after a tedious calculation, the short time expansion (i.e. small $\lambda$)
expansion of:
\bea
&& \overline{z^3}^c = 42.99376
~\lambda^3 \\
&& \overline{\ln z} = -\frac{3}{2} \sqrt{\frac{\pi}{2}} \lambda^{3/2}-0.27162097
~ \lambda ^3 \\
&& \overline{(\ln z)^2}^c = \frac{3}{2} \sqrt{2 \pi} \lambda ^{3/2}  +0.349154645
 ~  \lambda ^3 \nn  \\
&& \overline{(\ln z)^3}^c = 0.58226188
~ \lambda^3 
\eea 
up to $O(\lambda^{9/2})$ terms. The skewness of the PDF of $\ln z$ behaves at short time as:
\bea
\gamma_1 =  \frac{\overline{(\ln z)^3}^c}{(\overline{(\ln z)^2}^c)^{3/2}} \simeq 0.079863175
~ \lambda^{3/4}
\eea 
It is interesting to compare with the same results in Ref. \cite{we} in the absence of the hard wall (full space) 
and we find the universal ratio of the variances:
\be \label{R}
\rho = \frac{\overline{(\ln z)^2}^{c,HS}}{\overline{(\ln z)^2}^{c,FS}} = \frac{3}{2} - 0.076597089
~ \lambda
   ^{3/2}+O(\lambda ^{3})
\ee
and of the skewness:
\bea \label{rat}
\gamma_1^{HS}/\gamma_1^{FS} = 0.63689604
+ O(\lambda^{3/2})
\eea 
at small time. 

We now study arbitrary time, i.e. any $\lambda$, and to this aim 
we define the generating function of the distribution $P(f)$ of the 
scaled free energy $\ln Z =- \lambda f$:
\begin{eqnarray}  \label{defg0}
&& g(s) = \overline{ \exp(- e^{- \lambda s} Z) } = 1 + \sum_{n=1}^\infty \frac{(- e^{- \lambda s})^n}{n!} \overline{Z^n} 
\end{eqnarray} 
from which $P(f)$ is immediately extracted at $\lambda \to \infty$:
\begin{eqnarray}  
&& \lim_{\lambda \to \infty} g(s) = \overline{ \theta(f+s) } = Prob(f > -s) 
\end{eqnarray} 
and below we recall how it is extracted at finite $\lambda$. 
The constraint $\sum_{i=1}^{n_s} m_i=n$ in (\ref{partsum}) can then be relaxed
by reorganizing the series according to the number of strings:
\begin{eqnarray} \label{defg}
g(s) = 1 +  \sum_{n_s=1}^\infty \frac1{n_s!}  Z(n_s,s) 
\end{eqnarray}
Solvability for the generating function 
arises from the pfaffian identity:
\be \label{pfid}
 \prod_{1 \leq i<j \leq n_s} D_{k_i,m_i,k_j,m_j}  = \prod_{j=1}^{n_s} \frac{m_j}{2 i k_j} {\rm pf} ( \frac{X_i-X_j}{X_i+X_j} )_{2 n_s \times 2 n_s} 
\ee
where $X_{2p-1} = m_p + 2 i k_p$, $X_{2p} = m_p - 2 i k_p$, $p=1,..n_s$, 
a consequence of Schur's identity as used in Ref. \cite{we-flat,we-flatlong} to which we
refer for details. We recall that the pfaffian of an antisymmetric matrix $A$ is defined as ${\rm pf} A=\sqrt{\det A}$. Eq. (\ref{pfid}) allows to write
the $n_s$ string partition sum as \cite{footnote1}:
\bea
&& Z(n_s,s) = \sum_{m_1,..m_{n_s}=1}^\infty (-1)^{\sum_p m_p} \prod_{p=1}^{n_s}  
\int \frac{dk_p}{2 \pi} \frac{b_{m_p,k_p}}{4 i k_p} \nn \\
&& \times 
e^{ m_p^3 \frac{ t}{12} - m_p k_p^2 t - \lambda m_p s }  {\rm pf} ( \frac{X_i-X_j}{X_i+X_j} )_{2 n_s \times 2 n_s}
\eea
Now, as in Ref. \cite{we-flat,we-flatlong} we use the representation
$\int_{v_i,v_j>0} 2 \delta'(v_i-v_j) e^{-v_i X_i - v_j X_j} =  \frac{X_i-X_j}{X_i+X_j}$
and standard properties of the pfaffian allow to take the
integral over the $2 n_s$ variables outside the pfaffian. After 
manipulations very similar to Ref. \cite{we-flat,we-flatlong} the integration and summation
over $k_j,m_j$ can be performed, leading to:
\bea
Z(n_s,s) = \frac{1}{(2 n_s-1)!!} && \prod_{j=1}^{2 n_s} \int_{v_j>0} {\rm pf}( f(v_i,v_j) )_{2 n_s \times 2 n_s} \nn \\
&& \times {\rm pf} ( \delta'(v_i-v_j) )_{2 n_s \times 2 n_s} 
\eea
where $(2 n_s-1)!!=(2 n_s)!/(n_s! 2^{n_s})$ is the number of pairing of $2 n_s$ objects,
with the kernel:
\bea
 f(v_1,v_2) & = & \sum_{m=1}^\infty \int \frac{dk}{2 \pi}  \frac{(-1)^m b_{k,m}}{2 i k}  e^{ m^3 \frac{\lambda^3}{3} - 4 m k^2 \lambda^3 - \lambda m s }  \nn
\\
&& \times e^{- m( v_{1} + v_2) - 2 i k (v_1-v_2)} 
\eea 
where we used that in the natural units $t(\equiv \frac{t}{t^*})=4 \lambda^3$.
Hence $g(s)$ has now the form of a Fredholm Pfaffian and one shows \cite{us-long}:
\bea \label{resg}
&& g(s) = \sqrt{ {\rm Det}[ I + {\cal K}  ] } \\
&& {\cal K}(v_1,v_2) = - 2 \theta(v_1) \theta(v_2) \partial_{v_1} f(v_1,v_2) \nn
\eea
It is interesting that $g(s)^2$ is precisely the generating function for the two independent
half spaces (on each side of the hard wall) and that it is itself a Fredholm determinant (FD). Performing the rescaling
$v_j \to \lambda v_j$ and $k_j \to k_j/\lambda$ leaves the result (\ref{resg}) 
unchanged with the scaled kernel:
\bea 
&& f(v_1,v_2) = \int \frac{dk}{2 \pi} \int_y Ai(y+s+v_1+v_2+4 k^2) \nn \\
&& \times  f_{k/\lambda}(e^{\lambda y}) \frac{e^{- 2 i k (v_1-v_2)} }{2 i k} \label{frescaled}
\eea 
where we have used the now standard Airy trick $\int_y Ai(y) e^{w y} = e^{w^3/3}$
to transform the cubic exponential in an exponential, together with the
shift $y \to y+s+v_1+v_2$. The weight function $f_{k}(z) :=  \sum_{m=1}^\infty b_{k,m}  (-z)^m$
can be calculated explicitly and we find:
\bea \label{resfk} 
&& f_{k}[z] = \frac{2 \pi  k}{\sinh (4 \pi  k) } \left(J_{-4 i
   k}(\frac{2}{\sqrt{z}})+J_{4 i
   k}(\frac{2}{\sqrt{z}})\right) \nn \\
   && -\,
   _1F_2\left(1;1-2 i k,1+2 i k;- 1/z \right)
\eea 
Eqs. (\ref{resg}), (\ref{frescaled}) and (\ref{resfk}) is our main result at
finite time for $g(s)$.

We now obtain the PDF of the free energy (i.e. of the KPZ height), first at large time,
i.e. for $\lambda \to \infty$. Examination of (\ref{resfk}) leads to \cite{us-long}:
\bea
\lim_{\lambda \to \infty} f_{k/\lambda}[e^{\lambda y}]  = - \theta(y) (1 - \cos(2 k y))
\eea 
Rescaling $k \to k/2$ and taking the derivative in (\ref{resg})
one finds after integrations by part w.r.t. $y$:
\bea
&& g(s)^2 = {\rm Det}[ I + P_0 K_s P_0 ] = {\rm Det}[ I + P_{s/2} K_0 P_{s/2} ] \nn  \\
&& K_s(v_i,v_j) = - \int \frac{dk}{2 \pi} \int_{y>0} Ai(y + s+v_i+v_j+ k^2) \nn \\
&& \times e^{-  i (v_i-v_j) k} (1-e^{i k y}) 
\eea 
where $P_x \equiv \theta(v-x)$ projects all $v_j$ integrations on $[x,+\infty[$. 
Upon using the Airy function identity 
\bea
2^{-\frac{1}{3}} Ai(2^{\frac{1}{3}} a) Ai(2^{\frac{1}{3}} b) = \int \frac{dq}{2 \pi} Ai(q^2 + a + b) e^{i q (a-b)} \nn
\eea 
we find $K_0(v_i,v_j) = - 2^{1/3} \tilde K(2^{1/3} v_i, 2^{1/3} v_j)$ and 
upon rescaling of the $v_i$:
\bea \label{res1}
&& g(s)^2 = {\rm Det}[ I - P_{{\sf s}} \tilde K P_{{\sf s}} ] \quad , \quad {\sf s} =2^{-2/3} s\\
&& \tilde K(v_i,v_j) = K_{Ai}(v_i,v_j)  - \frac{1}{2} Ai(v_i) \int_{y>0} Ai(y+v_j)  \nn
\eea
where $K_{Ai}$ is the Airy Kernel $K_{Ai}(v_i,v_j) = \int_{y>0} Ai(v_1+y) Ai(v_2+y)$.
Our result (\ref{res1}) for the half space at large time can be compared 
with the full space result \cite{spohnKPZEdge,we,dotsenko,corwinDP}
$g_{FS}(s)={\rm Det}[ I - P_{{\sf s}} K_{Ai} P_{{\sf s}} ]  = F_2({\sf s})$, i.e. the
GUE distribution. Hence, as compared to the two half spaces, 
the second term (projector) in (\ref{res1}) encodes for the effect of the 
DP configurations which in full space, cross $x=0$ at least once. 
Interestingly since $Z_{FS} > Z_{HS}^{(1)} + Z_{HS}^{(2)}$ and the
two half spaces are statistically independent, one shows from the 
definition (\ref{defg0}) that:
\bea
g_{FS}(s) < g_{HS}(s)^2
\eea 
a bound valid at all times (not just for infinite $\lambda$). 

We can now transform our result (\ref{res1}) into a more familiar form. 
Defining ${\cal K}^{\infty}_{\sf s}(v_1,v_2)=\theta(v_1) \theta(v_2) \tilde K(v_1+{\sf s},v_2+{\sf s})$
we note that this Kernel can also be written as 
${\cal K}^{\infty}_{\sf s} = {\cal B}_{\sf s}^2 - \frac{1}{2} |{\cal B}_{\sf s} \delta \rangle \langle 1 {\cal B}_{\sf s}|$
where ${\cal B}_{\sf s}(x,y) := \theta(x) Ai(x+y+{\sf s}) \theta(y)$ 
one obtains via manipulations similar to Ref. 
\cite{we-flat,we-flatlong,ferrari2} in the infinite time limit:
\bea
&& g(s)^2 = Det[I - {\cal K}^{\infty}_{\sf s}] \\
&& = Det[I- {\cal B}_{\sf s}^2] (1 + \frac{1}{2} \langle 1 |\frac{{\cal B}_{\sf s}^2}{1-{\cal B}_{\sf s}^2}|\delta \rangle ) \\
&& = \frac{1}{4} (Det(I-{\cal B}_{\sf s}) + Det(I+{\cal B}_{\sf s}))^2 
\eea
using that $\langle 1 |\frac{1}{1\pm {\cal B}_{\sf s}}|\delta \rangle = Det(I \mp {\cal B}_{\sf s})/Det(I \pm {\cal B}_{\sf s})$.
Since $F_1({\sf s})=Det(I-{\cal B}_{\sf s})$, $F_2({\sf s})=Det(I-{\cal B}_{\sf s}^2)$
and using the definitions in \cite{baiksymptotics} we obtain \footnote{note that
other conventions for $F_4$ (e.g. wikipedia) differ by a factor $\sqrt{2}$}
\bea
&& g(s) = \frac{1}{2} (F_1({\sf s}) + \frac{F_2({\sf s})}{F_1({\sf s})}) = F_4({\sf s}=2^{-2/3} s) 
\eea 
Hence, to summarize, we find that for the continuum DP model in presence of the hard wall
one can write:
\bea \label{asymptz}
\ln z = 2^{2/3} \lambda ~ \xi_t  
\eea 
where $z=Z/\overline{Z}$ and $\xi_t$ converges at large time in distribution to the GSE 
Tracy Widom distribution $F_4$. The same formula holds for the full space 
but with $\xi_t$ converging at large time to the GUE distribution $F_2$.

We now obtain the PDF of the free energy at finite time. To this aim one follows the
method used in \cite{we}. It is written as a convolution, i.e. $\ln Z = \ln Z_0 + \lambda u_0$ is the sum of two independent
random variables, where $\ln Z_0$ has a unit Gumbel distribution
(i.e. $P(Z_0)=e^{-Z_0}$). Then the PDF of $u$ is obtained by analytical
continuation $p(u) = \frac{\lambda}{\pi}  {\rm Im} g(s)|_{e^{\lambda s} \to - e^{\lambda u} + i 0^+}$.
Using (\ref{resg}), (\ref{frescaled}) and (\ref{resfk}) and some complex analysis we
find the free energy distribution as the difference of two (complex) Fredholm
Pfaffians (FP):
\be \label{pu}
p(u) = \frac{\lambda}{2 i \pi} ( \sqrt{{\rm Det}[ I + P_{0} K_u P_{0} ]}  - \sqrt{{\rm Det}[ I + P_{0} K^*_u P_{0} ] } ) 
\ee
with the kernel:
\bea
&& K_u(v_i,v_j) = \frac{d}{d v_i} \int \frac{dk}{2 \pi} \int_{y} Ai(y +u+v_i+v_j+ 4 k^2)  \nn \\
&& \times \frac{\sin(2 (v_i-v_j) k)}{k}
[ f^r_{k/\lambda}(e^{\lambda y}) + i f^i_{k/\lambda}(e^{\lambda y}) ] \\
&& f_k^r(z) =  \frac{\pi  k}{\sinh(2 \pi  k)} (I_{-4 i k}(2 \sqrt{\frac{1}{z}})+I_{4 i k}(2 \sqrt{\frac{1}{z}})) \\
&& - 
   _1F_2\left(1;1-2 i k,1+2 i k;\frac{1}{z}\right) \\
   && f_k^i(z) =  4 k \sinh(2 k \pi) K_{4 i k}(2 \sqrt{\frac{1}{z}})
\eea 
Note that the same formula (\ref{pu}) with each FP replaced by its square, i.e. the FD, holds 
for the free energy associated to the union of the two independent half spaces.

{\it Numerical simulations} We now perform numerical checks. Here we call $\hat t$ the 
(integer) polymer length. At high temperature, we follow \cite{we,highT,flatnumerics} and
define the partition sum (PS) $Z(\hat t)=\sum_{\gamma_{\hat t}} e^{- \frac{1}{T} \sum_{(x,\tau) \in \gamma_{\hat t}} V(x,\tau)}$ 
of paths $\gamma_{\hat t}$ directed along the
diagonal of a square lattice from $(0,0)$ to $(\hat t/2,\hat t/2)$
with only $(1,0)$ or $(0,1)$ moves. We denote space 
$x=(i-j)/2$ and time $\tau=i+j$. An i.i.d. random number $V(x,\tau)$ is defined
at each site of the lattice (we use a unit centered Gaussian). The disorder averaged full space
PS is $\overline{Z}=N_{\hat t} e^{\beta^2 \hat t/2}$ where $N^{FS}_{\hat t} \simeq 2^{\hat t} \sqrt{2/(\pi \hat t)}$ is the number of
paths of length $\hat t$. The half space PS is obtained by summing 
only on paths with $x \geq 0$, in effect equivalent to an absorbing
wall (hard wall), with $N^{HS}_{\hat t} \simeq 2^{\hat t} (2/\hat t)^{3/2}/\sqrt{\pi}$.
We use the transfer matrix algorithm. It gives $\ln z$ as an output 
with $z=Z/\overline{Z}$. As was established in \cite{we,highT,flatnumerics}
in the high $T$ limit at fixed $\lambda$, where $\lambda= (\hat t/(2 T^4))^{1/3}$
for the lattice model, $\ln z$ can be directly compared - {\it with no free parameter} - 
with the analytical predictions of the continuum model with the same
value of $\lambda$, defined there by (\ref{lambdadef}). 
In addition we also perform numerics at
$T=0$ and compute the optimal path energy. 

\begin{figure}
\includegraphics[scale=0.27]{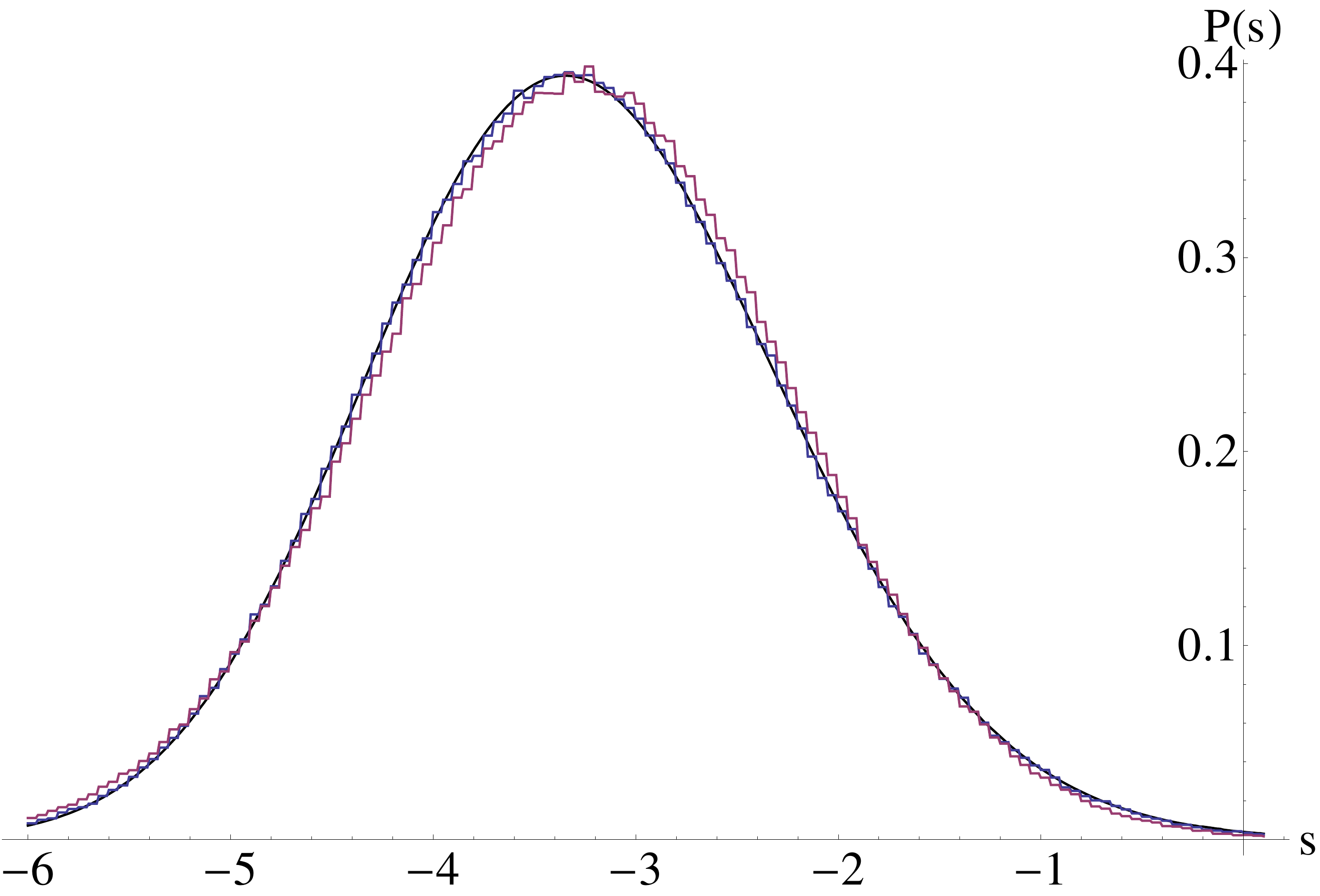}
\centering
\caption{Rescaled PDF of (minus) the free energy at large time. (i) solid line: analytical prediction $\frac{d}{d{\sf s}} F_4({\sf s})$. Histograms: (ii) \textit{in blue}, ground state energy PDF ($T=0$) for a polymer $\hat t=2^{10}$ with $N=10^6$ samples (iii) \textit{in red}, PDF of ${\sf s}=- 2^{-2/3}f$ for a polymer $\hat t=2^{10}$ at $\lambda=6.3$, with $N=10^6$ samples. The numerical PDFs are rescaled to adjust the mean and the variance of $F_4$.
The variable $s$ in all figures is called ${\sf s}$ in the text.}
\label{fpdf}
\end{figure}

\begin{figure}
\includegraphics[scale=0.3]{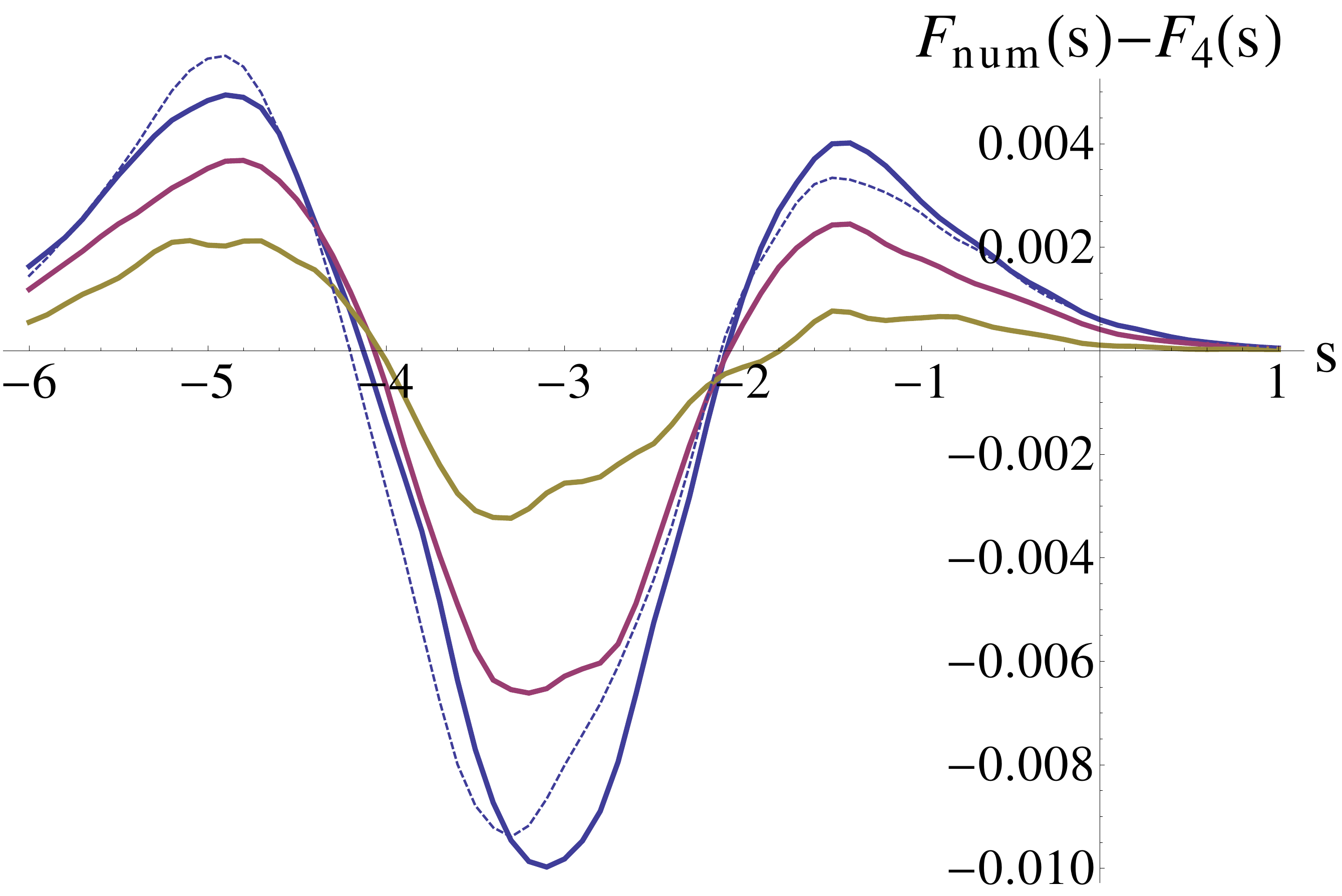}
\centering
\caption{Convergence as a function of $\lambda$: the difference between the numerical CDFs, $F_{num}(s)$, and the 
prediction for infinite $\lambda$, $F_4(s)$, is plotted for $\lambda = 0.2 \text{  (in blue)},1 \text{  (in red)},3 \text{  (in yellow)}$ with $N=2.10^5$ samples. $\hat t = 2^{11}$ is hold fixed. For $\lambda=0.2$ a length $\hat t=2^{9}$ is also shown (dashed line) illustrating finite size
effects. The statistical fluctuations due to finite sample $N$ are visible on the figure.}
\label{finitesizelambda}
\end{figure}

\begin{figure}
\includegraphics[scale=0.53]{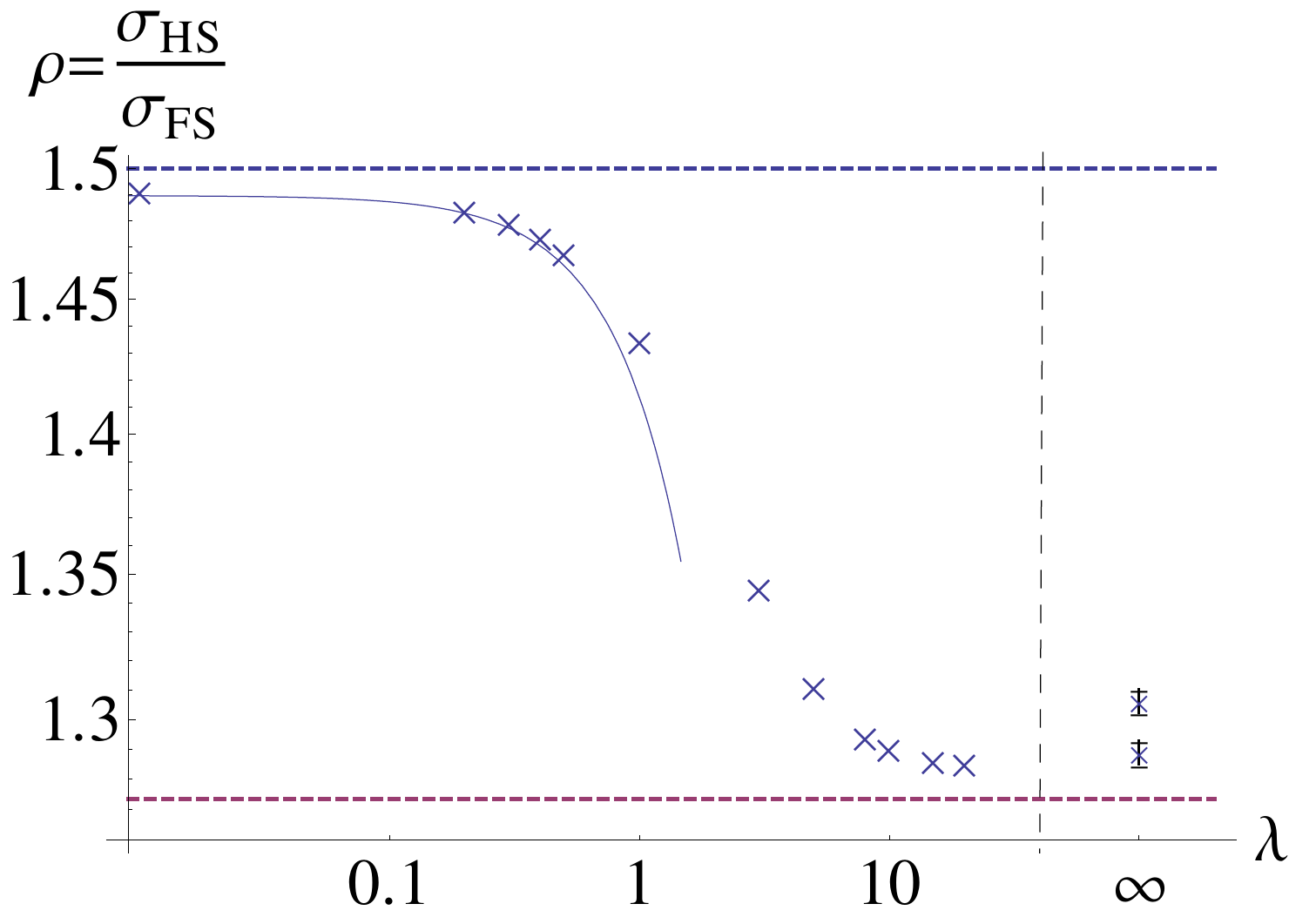}
\centering
\caption{Ratio of variances $\rho=\frac{\sigma _{HS}}{\sigma_{FS}}$, for $\lambda$ varying from $0.2$ to $20$. 
The crosses correspond to numerical data ($N =2\cdot 10^5$ samples, $\hat t=2^{11}$, standard error estimation $\epsilon=3\cdot 10^{-3}$). The dashed horizontal lines represent analytic predictions in both limits, $\frac{3}{2}$ for $\lambda \rightarrow 0$ and $\frac{\sigma^{F_4}}{\sigma^{F_2}}=1.2733$ for  $\lambda \rightarrow \infty$. The effect of finite $\hat t$ is clearly visible. It causes a small gap between these limits and the numerical data, which decreases as $\hat t$ increases. The solid line represents the Taylor expansion (\ref{R}) 
globally rescaled to account for finite $\hat t$. The right part of the graph shows the convergence of $\rho$ at $T=0$ as a function of $\hat t$: the upper point is $\hat t=2^8$, the lowest $\hat t=2^{10}$.}
\label{variancecrossover}
\end{figure}

In Fig. \ref{fpdf} we show the convergence to the GSE TW distribution
both for (i) $T=0$ and large polymer length $\hat t$ and (ii) at $T>0$
and large $\lambda$. The agreement is very good. The variation
for $T>0$ as a function of $\lambda$ is shown in more details 
in Fig. \ref{finitesizelambda} where the (small) differences in the
cumulative distributions (CDF) are shown on a larger scale. 
As in the previous figure the mean and variance of the numerical
PDF's are adjusted to those of $F_4$, hence this plot only shows variation
of the {\it shape} of the PDF. The variance and mean are
studied separately. In Fig. \ref{variancecrossover}
we show the ratio of half space (HS) to full space (FS)
variances as a function of $\lambda$. Since the two TW distributions 
have variances $\sigma^{F_2}=0.8131947928$ 
and $\sigma^{F_4}=1.03544744$, the ratio $\rho$ should converge to the value
$1.273308$ at large time, which is apparent in the
Fig. \ref{variancecrossover}, up to finite $\hat t$ effects discussed there.
Similarly the two TW distributions 
have skewness $\gamma_1^{F_2} = 0.2240842$ and
$\gamma_1^{F_4} = 0.16550949$ hence the skewness ratio 
is predicted to increase from $0.636896$ at small time, Eq. (\ref{rat}), 
to $0.738604$ at large time, a moderate variation. However
the combination of finite size effects and finite sample make it
hard to compute it with precision for small and for very large $\lambda$.
For $\lambda=5$ we find a value consistent with the above
variation interval. Finally the kurtosis are predicted to converge
at large time to $\gamma_2^{F_2}=0.093448$
and $\gamma_2^{F_4}=0.0491952$.

Interestingly, the difference of the means $\mu$ of the two TW distributions
(GSE and GUE) gives information about extreme value properties
of the DP. Let us define $p=Z^{HS}/Z^{FS}$ the probability that, in the full space problem and
for endpoints fixed at position $x>0$ the DP does not cross $x=0$. $p$ is defined for each
disorder realization, with   $p \simeq x^2 \tilde p/t$ for small $x$.
Then at large time (i.e. large $\lambda$) one has:
\bea
\overline{\ln \tilde p} =  2^{2/3} \lambda (\mu^{F_4} - \mu^{F_2})
\eea 
where $\mu^{F_4}=-3.2624279$ while $\mu^{F_2}=-1.7710868$. 
At small time (i.e. small $\lambda$) one finds from the above
results (and the ones in \cite{we}) 
$\overline{\ln \tilde p}  = -\frac{1}{2} \sqrt{\frac{\pi}{2}} \lambda^{3/2}-0.0082964 \lambda^3+..$
hence $- \overline{\ln \tilde p}$ crosses over from $\sim t^{1/2}$ to $\sim t^{1/3}$. 
Note that $p$ is highly non self-averaging at low temperature: at $T=0$
it is either 0 or 1, and a numerical study \cite{us-long} indicates that $\overline{p}={\rm Prob}(p=1)$ decays
algebraically with time. Computing the full distribution of $p$ seems a hard, although interesting, task.

%
{\it KPZ equation:} Let us now detail how our results translate in terms of 
the KPZ equation 
\bea
\partial_t h = \nu \nabla^2 h + \frac{\lambda_0}{2} (\nabla h)^2 + \eta(x,t)
\eea 
where $\overline{\eta(x,t) \eta(x',t')}=R_\eta(x-x') \delta(t-t')$, with Gaussian noise correlator $R_\eta(x) = D \delta(x)$. The Cole-Hopf
mapping generally implies:
\bea
\frac{\lambda_0}{2 \nu} h = \ln Z \quad , \quad \bar c = D \lambda_0^2
\eea 
Here however we must be more specific. The initial condition (\ref{zdef}) corresponds to 
a wedge $h(x,0)=- w|x-y|$ in the limit $w \to \infty$, before $y \to 0$. Because of
the hard wall we have that $\frac{\lambda_0}{2 \nu} h(x,t) = \ln(x y) + \frac{\lambda_0}{2 \nu}  \tilde h(x,t)$
where $\tilde h$ is not singular when both $x$ and $y$ approach zero,
and the correspondence is really $\frac{\lambda_0}{2 \nu} \tilde h(0,t) = \ln Z$.
Schematically the boundary conditions (BC) can be stated as 
$h(0,t)=-\infty$ or $\nabla h(0,t)=+\infty$ 
(see more general ones below).
Hence from (\ref{asymptz}):
\bea
\frac{\lambda_0}{2 \nu} \tilde h(0,t) = \ln \overline{Z} +  2^{2/3} \lambda ~ \xi_t  
\eea 
with, at large $t$, $\ln \overline{Z} \simeq v_\infty t$. From \cite{footnote0,footnote1},
$v_\infty = \frac{\lambda_0^2 R_\eta(0)}{8 \nu^2}-\frac{D^2 \lambda_0^4}{12}$ is the
same non universal constant (see discussion in \cite{flatnumerics}) in both HS and FS cases, the
difference in $\ln \overline{Z}$ being only sublinear in time, as $\sim \ln t$.

Finally, we discuss the universality of our results. The BC we used here at $x=0$ is the
hard wall $Z=0$, which in the KPZ context corresponds to $\nabla h=+\infty$. 
Another standard BC is the reflecting wall (RW) $\nabla Z=0$, i.e. $\nabla h=0$ (contact angle $\pi/2$). For the DP it
can be achieved by considering two symmetric half-spaces i.e.
$V(-x,t)=V(x,t)$ \cite{footnote2}. At $T=0$ there is no difference in the optimal path energy between 
the hard and reflecting wall, see Fig. \ref{f:dp}. At $T>0$ the two cases become different, since
there is more entropy in the RW. However the longer the polymer, 
the closer it becomes, effectively, to the zero temperature limit. Hence we expect that
although at finite time the two cases lead to different $g(s)$, these become
equal at large time. In fact all BC such that $\nabla h \geq 0$ should converge
to $F_4$. This is consistent with the results of \cite{BaikSymPermutations}
translated into the $T=0$ lattice DP model (although the equivalent of the hard wall was not explicitly considered there). 
In the PNG model it corresponds to the absence of boundary source,
or a weak enough source \cite{sasamotohalfspace}. We will not discuss here the case of BC $\nabla h < 0$ 
which leads to an unbinding transition. A similar transition was studied in the random permutation
model \cite{BaikSymPermutations} and in the PNG model \cite{spohn2000,sasamotohalfspace}, but not 
using the BA (see however \cite{kardarunbinding}).
Work on that case is in progress. 

It is worth pointing out an application of our results to the conductance
$g$ of disordered 2D conductors deep in the localized regime. Extending the results of Ref.
\cite{ortuno} we predict that $L^{-1/3} \ln g$ should be distributed as $F_4$ if the leads are small, separated by $L$, and 
placed near the frontier of the sample 
(which occupy, say, a half space).

We thank P. Calabrese for numerous discussions and
pointing out Ref. \cite{BAhardwall}. We thank A. Rosso
for helpful remarks. We are grateful to N. Crampe, A. Dobrinevski
and M. Kardar for interesting discussions and 
pointing out Ref. \cite{kardarunbinding}. 
This work was supported by ANR grant 09-BLAN-0097-01/2.


\begin{thebibliography}{100}
\bibitem{footnote0}
One has $\overline{Z}=1/\sqrt{4 \pi t}$ for full space
and $\overline{Z}=1/\sqrt{4 \pi} t^{3/2}$ with the hard wall, as in 
absence of disorder. The non-universal global multiplicative constant
$e^{R_V(0) t/(2 T^2)}$ does not affect the variable $z$ and has been dropped.
\cite{flatnumerics}.
\bibitem{footnote1}
Here we have performed the usual shift $Z = e^{-c^2 t/12} \hat Z$ (we drop the hat below)
which does not affect the variable $z=Z/\overline{Z}$.
\bibitem{footnote2}
Note that if one chooses $V(-x,t)=V(x,t)$ the usual image method works
i.e. $Z(x,y,t) \pm Z(x,-y,t)$ is the PS in the half space with reflecting (resp.
absorbing) BC.
\end{thebibliography}
\end{document}